\documentclass[twocolumn]{aa}
\usepackage{graphicx}
\usepackage{txfonts}
\begin{document}
   \title{A Giant Planet Candidate near a Young Brown Dwarf \thanks{Based on observations obtained at the Paranal Observatory, Chile, in ESO programs 73.C-0469 and 273.C-5029}}

   \subtitle{Direct VLT/NACO Observations using IR Wavefront Sensing}

\author{
        G. Chauvin\inst{1}\and
        A.-M. Lagrange\inst{2}\and
        C. Dumas\inst{1}\and
        B. Zuckerman\inst{3}\and
        D. Mouillet\inst{4}\and
	I. Song\inst{3}\and
        J.-L. Beuzit\inst{2}\and
        P. Lowrance\inst{5}
}

 \offprints{Ga\"el Chauvin \email{gchauvin@eso.org}}
\institute{
$^{1}$European Southern Observatory, Casilla 19001, Santiago 19, Chile\\
$^{2}$Laboratoire d'Astrophysique, Observatoire de Grenoble, 414, Rue de la piscine, Saint-Martin d'H\`eres, France\\
$^{3}$Department of Physics \& Astronomy and Center for Astrobiology, University of California, Los Angeles, 8371 Math Science Building, Box 951562, CA 90095-1562, USA\\             
$^{4}$Laboratoire d'Astrophysique, Observatoire Midi-Pyr\'en\'ees, Tarbes, France\\
$^{5}$Spitzer Science Center, Infrared Processing and Analysis Center, MS 220-6, Pasadena, CA 91125, USA\\
}

   \date{Received September 15, 1996; accepted March 16, 1997}

   \abstract{
We present deep VLT/NACO infrared imaging and spectroscopic observations of the brown dwarf 2MASSWJ\,1207334$-$393254, obtained during our on-going adaptive optics survey of southern young, nearby associations. This $\sim25$~M$_{\rm{Jup}}$ brown dwarf, located $\sim70$\,pc from Earth, has been recently identified as a member of the TW Hydrae Association (age $\sim8$~Myr). Using adaptive optics infrared wavefront sensing to acquire sharp images of its circumstellar environment, we discovered a very faint and very red object at a close separation of $\sim780$~mas ($\sim55$ AU). Photometry in the H, $\rm{K}_{s}$ and $\rm{L}\,\!'$ bands and upper limit in J-band are compatible with a spectral type L5-L9.5. Near-infrared spectroscopy is consistent with this spectral type estimate. Different evolutionary models predict an object within the planetary regime with a mass of $\rm{M}=5\pm2$~M$_{\rm{Jup}}$ and an effective temperature of $\rm{T}_{\rm{eff}}=1250\pm200$~K. 

   \keywords{2MASSWJ\,1207334$-$393254 -- brown dwarf -- giant planet  -- adaptive optics imaging and spectroscopy}
   }

   \maketitle
%

\section{Introduction}
 
In recent years, the field of extrasolar planet detection and characterization has been exclusively reserved to the domain of indirect detection measurements, radial velocity surveys (Fisher et al. 2003, Mayor et al. 2004) or transit detections (Charbonneau et al. 2000, Konacki et al. 2004). However, this exploration is presently intrinsically limited to the close circumstellar environment within $\sim4$~AU. With the recent development of high contrast and high angular resolution instrumentation, the situation is about to change and the exploration of planets with large semi-major axes is now achievable.

In this letter, we report the discovery of a probable giant planet companion from direct imaging techniques. In the course of our on-going deep imaging survey of young, nearby southern associations (Chauvin et al. 2003), we used the ESO VLT telescope and its adaptive optics near-infrared instrument NACO to image the close vicinity of the source 2MASSWJ\,1207334$-$393254 (here after 2M1207). This brown dwarf 2M1207 was identified by Gizis (2002) as a member of the young TW~Hydrae Association (TWA). We present here the observations and the reduction techniques used, as well as the infrared photometric, astrometric and spectroscopic results for 2M1207 and its giant planet candidate companion (GPCC). The likelihood of companionship, the age and the distance of the system and the mass predicted from several evolutionary models are finally discussed.

\begin{figure}
\centering
\includegraphics[height=6.5cm]{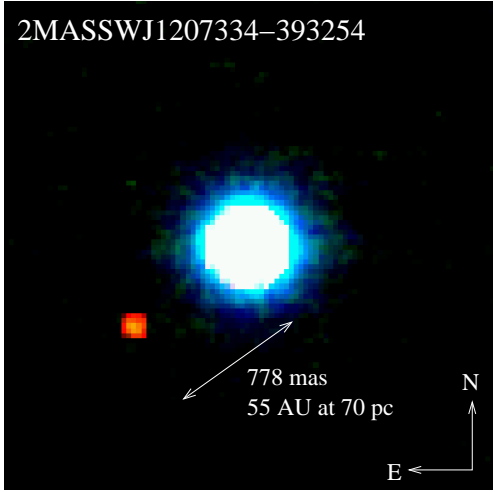}
     \caption{Composite image of brown dwarf 2M1207 and its GPCC in H (\textit{blue}), K$_{s}$ (\textit{green}) and L$\,\!'$ (\textit{red}). The companion appears clearly distinguishable in comparison to the color of the brown dwarf 2M1207.}
\label{fig:1}
\end{figure}


\section{Observations, reduction and results}

On 27 April 2004, we imaged the young brown dwarf 2M1207 with the NACO adaptive optics instrument of the VLT-UT4 (Rousset et al. 2002, Lenzen et al. 1998). Here, we used the unique capability offered by NACO to sense the wavefront in the near-infrared with the N90C10 dichroic ($90\%$ of the flux transmitted to the wavefront sensor and $10\%$ to the near-infrared camera CONICA). This mode is dedicated to the sharp imaging of red sources $\rm{V}-\rm{K}\ge6$~(M5 or later spectral type). 

\begin{figure}[t]
\centering
\includegraphics[height=6.5cm]{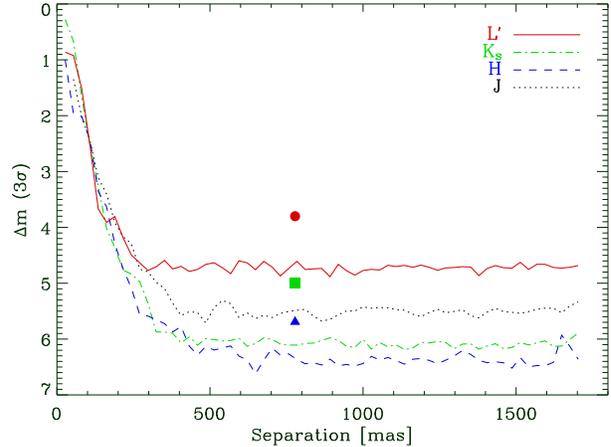}
     \caption{Detection limits at $3\sigma$ achieved during our observations in J-band (\textit{dotted black line}), H-band (\textit{dashed blue line}), K$_{s}$-band (\textit{dashed-dotted green line}) and L$\,\!'$-band (\textit{solid red line}). The contrasts between 2M1207 and its GPCC are reported for H (\textit{filled triangle}), K$_{s}$ (\textit{filled box}) and L$\,\!'$ (\textit{filled circle}) (the GPCC was not detected in J band).}
\label{fig:2}
\end{figure}

The source 2M1207 was then imaged in J, H, K$_{s}$ and L$\,\!'$ bands. The corresponding Strehl ratios, the full widths at half maximum intensity and other observing parameters in each band are given in Table~1. The AO IR sensing allowed us to close the adaptive optics loop on 2M1207 and to detect in its close vicinity a faint and red object at 778~mas and a position angle of $125.8^o$ in H, K and L$\,\!'$. The faint object was not detected down to 3$\sigma$ of 18.5 in J-band. In Fig.~\ref{fig:1} and~\ref{fig:2}, we display an H, K$_{s}$ and L$\,\!'$ composite image and the detection limits obtained in each band during our observations. After cosmetic reductions using  \textit{eclipse} (Devillar 1997), we used the myopic deconvolution algorithm \textit{MISTRAL} (Conan et al. 2000) to obtain H, K and L$\,\!'$ photometry and astrometry of the GPCC. The results are reported in Table~2. The transformations between the filters K$_{s}$ and K were found to be smaller than the measuring errors. 
\begin{table}[b]
\label{table:1}      
\begin{center}     
\caption{Night Log of the observations. S27 and L27 correspond respectively to a platescale of 27.03 and 27.12~mas. DIT and NDIT correspond respectively to an individual integration time and the number of integrations. Sr and FWHM correspond to the strehl ratio and the full width at half maximum intensity.}
\begin{tabular}{llllllll}
\hline\hline
  Filt.  &  Obj.  &  DIT  &  NDIT  &  Seeing   &   Airm.  &   Sr   &   \footnotesize{FWHM}                \\
   &          &  (s)  &        &  (\,$\!''$) &        &(\%)        &    (mas)         \\
\hline
\multicolumn{8}{c}{Imaging}\\
\hline
J         &   S27    & 30    & 8    & 0.59      & 1.07           & 6            & 122\\
H         &   S27    & 30    & 16   & 0.46      & 1.10           & 15           & 91\\
K$_s$   &   S27    & 30    & 16   & 0.52      & 1.08           & 23           & 89\\
L$\,\!'$   &   L27    & 0.175 & 1300 & 0.43      & 1.14           & 30           & 107\\
\hline
\multicolumn{8}{c}{Spectroscopy}\\
\hline
SH   &   S54    & 300 & 6 & 0.45      & 1.15           &            & \\
\hline
\end{tabular}
\end{center}
\end{table}

On 19 June 2004, 2M1207 and its GPCC were simultaneously observed using the NACO spectroscopic mode. The low resolution ($\rm{R}_{\lambda}=700$) grism was used with the 86~mas slit, the S54 camera (54~mas/pixel) and the SH filter (1.37-1.84~$\mu$m). The spectra of 2M1207 and its GPCC were extracted and calibrated in wavelength with \textit{IRAF/DOSLIT}. To calibrate the relative throughput of the atmosphere and the instrument, we divided the extracted spectra by the spectra of a standard star (HIP\,062522, B9III) and then multiplied by a blackbody to restore the shape of the continuum.

%

\section{Discussion}

\begin{table*}
\label{table:2}      
\begin{center}     
\caption{Photometric and astrometric measurements of 2M1207 and its GPCC obtained with the myopic deconvolution algorithm \textit{MISTRAL} (Conan et al. 2000). The transformations between the two filters K$_{s}$ and K were found smaller than the measuring errors.}
\begin{tabular}{lllllllllll}
\hline\hline
Source		         &SpT 	&Age	&J       &   H   &   K   & L$\,\!'$  &    $\Delta$	   & PA  \\
                         &	&(Myr)	&(mag)	 &(mag)  & (mag) & (mag)   & (mas) &  ($^o$)\\
\hline
2MASSWJ1207334$-$393254    &M8    &8$^{+4}_{-3}$	& $13.00\pm0.03\,^a$	 & $12.39\pm0.03\,^a$	 & $11.95\pm0.03\,^a$  &$11.38\pm0.10\,^b$   &     & \\
giant planet candidate   &L5-L9.5	&               & $\ge18.5\,^a$              & $18.09\pm0.21\,^a$    & $16.93\pm0.11\,^a$  &$15.28\pm0.14\,^b$   &778 & 125.8   \\
\hline 
\end{tabular}
\begin{list}{}{}
\item[$^{\mathrm{a}}$] from the 2MASS All-Sky Catalog of Point Sources (Cutri et al. 2003) and NACO contrast measurements (this paper)
\item[$^{\mathrm{b}}$] from L$\,\!'$ Keck I photometry (Jayawardhana et al. 2003) and NACO contrast measurements (this paper)
\end{list}
\end{center}
\end{table*}

\subsection{Membership in the TW Hydrae Association}

Gizis (2002) undertook a 2MASS-based search for isolated low mass brown dwarfs in the area covered by stellar members of TWA and found two late M-type objects which he identified as brown dwarfs.  The one of interest in the present paper, 2M1207, showed impressively strong H$\alpha$ emission in addition to signs of low surface gravity, which both are characteristic of very young objects. Gizis (2002) noted also that the proper motion of 2M1207 is consistent with membership in the TWA. 

Subsequently, Mohanty et al (2003) obtained echelle spectra of 2M1207. The radial velocity is also consistent with TWA membership. They detected a narrow Na\,\footnotesize{I}\normalsize~(8200\,$\AA$) absorption line indicating low surface gravity.  Finally, the spectrum displays various He\,\footnotesize{I}\normalsize~and H\,\footnotesize{I}\normalsize~emission lines (Mohanty et al 2003; Gizis 2002) and the H$\alpha$ line is asymmetric and broad.  Taken together, these characteristics led Mohanty et al (2003) to suggest the occurrence of ongoing accretion onto (a young) brown dwarf. Although $\rm{L}\,\!'$-band observations of Jayawardhana et al. (2003) did not reveal significant IR excess at 3.8~$\mu$m, recent mid-IR observations of Sterzik et al. (2004, accepted) found excess emission at 8.7~$\mu$m and 10.4~$\mu$m and confirm disk accretion as the likely cause of the strong activity. New \textit{Chandra} observations of Gizis \& Bharat (2004) corroborates this disk-accretion scenario as they suggest that less than 20\% of the H$\alpha$ emission can be due to chromospheric activity. All in all, multiple lines of evidence point toward membership of 2M1207 in the TWA.

\subsection{Age and distance of the system}

The age of the TWA can be established by comparison with the somewhat older $\beta$ Pictoris moving group's space motions (UVW; Zuckerman et al 2001) and HR diagrams. Ortega et al (2002) and Song et al (2003) have traced members back to a common volume 12 Myr ago which they identify as the age of the $\beta$ Pictoris moving group. On a color-magnitude diagram (Song et al 2003; Zuckerman \& Song 2004), the TWA stars lie slightly above the $\beta$ Pictoris stars as well as near 10~Myr isochrones derived by four different research groups. Consequently, an age of $8^{+4}_{-3}$ Myr is estimated for TWA. 

The spectral type of 2M1207 is estimated to be $\sim$M8 based on its broadband colors and spectra. An absolute K-magnitude of an M8 star is $\sim10.2$ (see Fig.~4 of Vrba et al. 2004 for example). However, young brown dwarfs are significantly brighter than their older counterparts. A K-band brightness difference between young ($\sim10$~Myr) and old late M-type stars can be estimated and 2.5 mag is obtained by extrapolating trends of young and old mid-M stars in a $\rm{V}-\rm{K}$ vs. M$_{\rm{K}}$ diagram of Song et al. (2003) and Zuckerman \& Song (2004). Then, a K-band distance modulus of $\sim4.3$ mag implies a distance $\sim70$~pc to 2M1207. Sterzik et al (2004) also appear to favor this distance.

\subsection{Spectral characterization}

\begin{figure}[t]
\centering
\includegraphics[height=8.5cm]{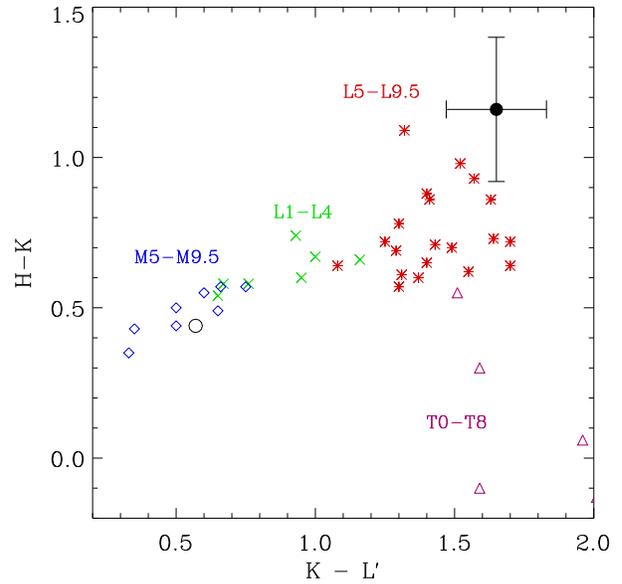}
     \caption{Color-color diagram ($\rm{H}-\rm{K}$ vs $\rm{K}-\rm{L}\,\!'$) for 2M1207 (\textit{open circle}) and its faint GPCC (\textit{point shown with error bars}). The observed infrared photometry for late-M dwarfs (\textit{diamonds}), L1-L4 dwarfs (\textit{crosses}), L5-L9.5 dwarfs (\textit{asterisks}) and T dwarfs (\textit{triangles}) obtained by Leggett et al. (2002) and Stephens et al. (2001) are given for direct comparison.}
\label{fig:3}
\end{figure}
By comparison of the GPCC photometry with the infrared colors of field late-M, L and T dwarfs obtained by Leggett et al. (2001, 2002), Stephens et al. (2001), Golimowski et al. (2004) and Knapp et al. (2004), we obtain a spectral type L5-L9.5 (see the color-color diagram of Fig.~3). 

Based on our NACO spectra recorded on 19 June 2004 (see Fig.~4), spectral types of 2M1207 and its GPCC can be derived by comparison with the observations of late-M, L and T dwarfs, published by Geballe et al. (2002) and Leggett et al. (2000, 2001). In the case of 2M1207, we find a spectral type M$8.5\pm1$ consistent with the estimation of Gizis (2002). For the GPCC, broad water-band absorptions are clearly distinguishable in the spectrum despite a low signal to noise (from 3 to 7$\sigma$). This allows us to derive a spectral type L5-L9.5, which confirms the estimation from IR colors. 

\subsection{Likelihood of a bound companion}

Although a spectral type L5-L9.5 confirms a substellar status for the GPCC, it does not allow us to confirm its companionship. Contamination by a foreground or background field L dwarf is still possible. The GPCC IR colors are consistent with that observed by Leggett et al. (2002) for late L field dwarfs, age of 1--5 Gyr; for example, an  L8 field dwarf located at $\sim60$~pc or L5 field dwarf at $\sim130$~pc. 

Based on a number density of L dwarfs of $1.9\times10^{-3}$~pc$^{-3}$ given by Burgasser (2001) and Cruz et al. (2003), the probability of finding a foreground or background L dwarf (located between 50 and 150~pc) in a region of 780~mas radius around 2M1207 is equal to~$9\times10^{-8}$. Consequently, contamination by an L field dwarf is very improbable.

\begin{figure}[t]
\centering
\includegraphics[height=8.5cm]{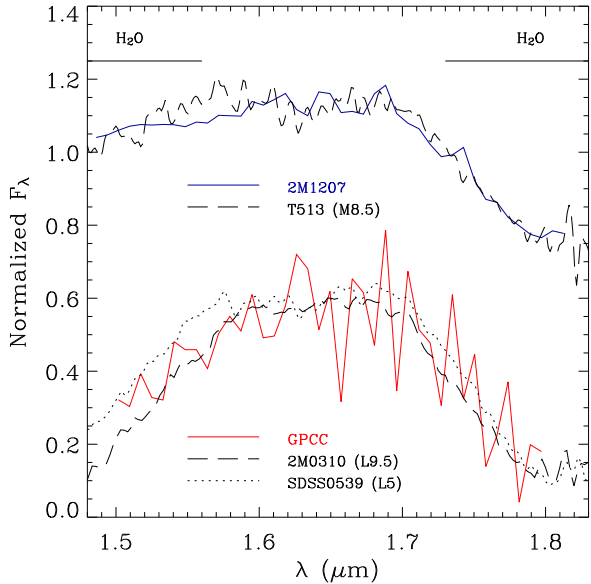}
     \caption{SH-band spectra of 2M1207 and its GPCC with the low resolution ($\rm{R}_{\lambda}=700$) grism of NACO, the 86~mas slit and the S54 camera (54~mas/pixel). The best adjustements were obtained with the template spectra of T513 (M8.5) for 2M1207, and SDSS0539 (L5) and 2M0310 (L9.5) for its GPCC.}
\label{fig:4}
\end{figure}

\subsection{Temperature and mass predictions}

In the case of a bound companion, evolutionary models can be compared to the photometry of the GPCC to derive the mass and the effective temperature expected. However, although the models are reliable for objects with age $\ge100$~Myr, they are more uncertain at early phases of evolution ($\le100$~Myr). As described by Baraffe et al. (2002), the choice of the initial conditions for the model adds an important source of uncertainty which is probably larger than the uncertainties associated with the age and distance of 2M1207.

If we consider the non gray model of Burrows et al. (1997), for a distance of 70~pc and age of $\sim5$--$10$~Myr, we derive a mass of 3--10~M$_{\rm{Jup}}$ and an effective temperature of 1000--1600~K. We then consider the new generation of models developed by Chabrier et al. (2000) and Baraffe et al. (2002). The DUSTY and COND models can be seen as two extreme cases, to describe respectively brown dwarfs with an atmosphere saturated in dust, like late M and L dwarfs, and cool brown dwarfs or giant planets with the dust condensed in their atmosphere, like T dwarfs. Based on the different color-magnitude diagrams using H, K and L$\,\!'$ bands or using the measured bolometric corrections BC$_{\rm{K}}\sim3.2$ for late-L dwarfs estimated by Golimowski et al. (2004), we derive a mass of $\rm{M}=5\pm2$~M$_{\rm{Jup}}$ and an effective temperature of $\rm{T}_{\rm{eff}}=1250\pm200$~K. This would give a mass ratio of 0.2 between the GPCC and the young brown dwarf 2M1207 with a projected separation of 55~AU.

%
%

\section{Conclusions}

We present high contrast observations of the young brown dwarf 2MASSWJ\,1207334$-$393254, a probable member of the TW Hydrae Association, obtained with adaptive optics and infrared wavefront sensing at the ESO VLT. We discovered a faint and extremely red object at $\sim780$~mas (55~AU) from this source. Photometry and spectroscopy are consistent with a spectral type L5-L9.5. For an age of $8^{+4}_{-3}$~Myr, evolutionary models return characteristics within the planetary regime: a mass of $\rm{M}=5\pm2$~M$_{\rm{Jup}}$ and an effective temperature of $\rm{T}_{\rm{eff}}=1250\pm200$~K. If confirmed, this object could be the first exo-planet companion ever imaged.

%
%

\begin{acknowledgements}
We would like to thank the staff of the ESO Paranal observatory, particularly Nancy Ageorges who first saw the GPCC. We thank also Gilles Chabrier, Isabelle Baraffe and France Allard for providing the latest update of their evolutionary models. Finally, we thank Sandy Leggett and Tom Geballe who kindly sent the near-infrared template spectra of M and L dwarfs and John Gizis for his remarks on the letter. 
\end{acknowledgements}


\begin{thebibliography}{}
%
\bibitem[Baraffe et al.(2002)]{bara02} Baraffe I., Chabrier G., Allard F. \& Hauschildt P.H., 2002, A\&A 382, 563
%
\bibitem[Burgasser(2001)]{burg01} Burgasser, A. 2001, Caltech PhD Thesis  
%
\bibitem[Burrows et al.(1997)]{burr97} Burrows, A., Marley, M., Hubbard, W. B. et al. 1997, AJ, 491, 856 
%
\bibitem[Chabrier et al.(2000)]{chab00} Chabrier, G., Baraffe, I., Allard, F. \& Hauschildt, P.H. 2000, ApJ, 542, 464
%
\bibitem[Chauvin et al.(2003)]{chau03} Chauvin, G., Thomson, M., Dumas, C. et al. 2003, A\&A, 404, 157
%
\bibitem[Charbonneau et al.(2000)]{char00} Charbonneau, D., Brown, T. M., Latham, D. W. \& Mayor, M. 2000, ApJ, 529, L45
%
\bibitem[2000]{cona00} Conan, J.-M., Fusco, T., Mugnier, L. et al. 2000, SPIE, Vol. 4007
%
\bibitem[Cruz et al.(2003)]{cruz03} Cruz, K.L., Reid, N.I., Liebert, J., Kirkpatrick, J.D. \& Lowrance P.J.  2003, AJ, 126, 2421
%
\bibitem[Cutri et al.(2003)]{cutr03} Cutri, R. M., Skrutskie, M. F., van Dyk, S. et al. 2003, 2MASS All-Sky Catalog of Point Sources
%
\bibitem[Devillar(1997)]{devi97} Devillar N. 1997, The messenger, 87
%
\bibitem[Fisher et al.(2003)]{fish03} Fischer, D. A., Butler, R. P., Marcy, G. W., Vogt, S. S. \& Henry, G. W. 2003, ApJ, 590, 1081
%
\bibitem[Geballe et al.(2002)]{geba02} Geballe, T. R., Knapp, G. R., Leggett, S. K. et al. 2002, ApJ, 564, 466
%
\bibitem[Gizis (2002)]{gizi02} Gizis, J.E. 2002, ApJ, 575, 484
%
\bibitem[Gizis (2004)]{gizi04} Gizis, J.E. \& Bharat, R. 2004, astroph-0405153
%
\bibitem[Golimowski et al.(2004)]{goli04} Golimowski, D. A., Leggett, S. K., Marley, M. S. et al. 2004, AJ, 127, 3516
%
\bibitem[Jayawardhana et al.(2003)]{jaya03} Jayawardhana, R., Ardila, D. R., Stelzer, B. \& Haisch, K. E. 2003, AJ, 126, 1515
%
\bibitem[Knapp et al.(2004)]{kanp04} Knapp, G.R., Leggett, S. K., Fan, X. et al. 2004, AJ, 127, 3553
%
\bibitem[Konacki et al.(2004)]{kona04} Konacki, M., Torres, G., Sasselov, D. D. et al. 2004, ApJ, 609, L37
%
\bibitem[Leggett et al.(2001)]{legg01} Leggett, S. K., Allard, F., Geballe, T. R., Hauschildt, P. H. \& Schweitzer, A. 2001, ApJ, 548, 908
%
\bibitem[Leggett et al.(2002)]{legg02} Leggett, S. K., Golimowski, D. A., Fan, X. et al. 2002, ApJ, 564, 452
%
\bibitem[Lenzen et al.(1998)]{lenz98} Lenzen, R., Hofmann, R., Bizenberger, P. \& Tusche, A., 1998, SPIE, Vol. 3354
%
\bibitem[Mayor et al. (2004)]{mayo04} Mayor, M., Udry, S., Naef, D. et al. 2004, A\&A, 415, 391 
%
\bibitem[Mohanty et al. (2003)]{moha03} Mohanty, S., Jayawardhana, R., Barrado y Navascu\'es, D. 2003, A\&A, 411, 517
%
\bibitem[Ortega et al.(2002)]{orte02}Ortega, V. G., de la Reza, R., Jilinski, E. \& Bazzanella, B. 2002, ApJ, 575, 75	
%
\bibitem[Rousset et al.(2002)]{rous02} Rousset, G., Lacombe, F., Puget, et al., 2002, SPIE, Vol. 4007
%
\bibitem[Song et al.(2003)]{song03} Song, I., Zuckerman, B. \& Bessell, M. S. 2003, ApJ, 599, 342
%
\bibitem[Stephens et al.(2001)]{steph01} Stephens, D. C., Marley, M. S., Noll, K. S. \& Chanover, N. 2001, ApJ, 556, 97
%
\bibitem[Sterzik et al.(2004)]{ster04} Sterzik, M.F., Pascucci, I., Apai, D., van der Blieck, N. \& Dullemond, C.P. 2004, astroph-0406460
%
\bibitem[Vrba et al.(2004)]{udry03} Vrba, F.J. et al. 2004, AJ, 127, 2948
%
\bibitem[Zuckerman et al.(2001)]{zuck01} Zuckerman, B.; Song, I.; Bessell, M. S. \& Webb, R. A. 2001, ApJ, 562, 87
%
\bibitem[Zuckerman \& Song(2004)]{zuck04}Zuckerman, B. \& Song, I. 2004, ARAA, v42, pp685-721

\end{thebibliography}
\end{document}